# Thermal and Athermal Swarms of Self-Propelled Particles

Nguyen Nguyen[1], Eric Jankowski[2] and Sharon C. Glotzer[2,3,*]

*Departments of Mechanical Engineering[1] Chemical Engineering[2], and Materials Science & Engineering[3], University of Michigan, Ann Arbor, MI 48109-2136, USA*

(Dated: December 30, 2011)

Abstract

Swarms of self-propelled particles exhibit complex behavior that can arise from simple models, with large changes in swarm behavior resulting from small changes in model parameters. We investigate the steady-state swarms formed by self-propelled Morse particles in three dimensions using molecular dynamics simulations optimized for GPUs. We find a variety of swarms of different overall shape assemble spontaneously and that for certain Morse potential parameters coexisting structures are observed. We report a rich "phase diagram" of athermal swarm structures observed across a broad range of interaction parameters. Unlike the structures formed in equilibrium self-assembly, we find that the probability of forming a self-propelled swarm can be biased by the choice of initial conditions. We investigate how thermal noise influences swarm formation and demonstrate ways it can be exploited to reconfigure one swarm into another. Our findings validate and extend previous observations of self-propelled Morse swarms and highlight open questions for predictive theories of nonequilibrium self-assembly.

---

,*Corresponding author: sglotzer@umich.edu

I. INTRODUCTION

The emergence of ordered swarms from a collection of autonomous self-propelled agents is a ubiquitous natural phenomenon. Flocks of birds, schools of fish, and herds of buffalo have been described by simple models that have two features in common: the agents that comprise the swarm should be driven towards an optimal velocity and there is an attraction between neighbors [1–4]. Swarms formed from bio-inspired interactions in which birds interact via three behavioral zones [5] are qualitatively similar to swarms formed by propelled point particles interacting via pairwise potentials [6, 7]. Analogues of the phase transitions observed in equilibrium systems are demonstrated in propelled systems [8–15], and recent work has elucidated different swarm structures as a function of interaction type [5–8,16–18]. Models including hydrodynamic interactions have also shown interesting collective behavior for a number of different active particles [3,19-22]. Exploiting the swarms formed in self-propelled systems has applications including control of unmanned aerial vehicles [23], assembly of mobile networks [24], and microscale mixing [25].

The generalized Morse potential, first introduced for modeling swarms by Levine *et al.* [6], has received considerable attention as a model of interparticle interactions because of its simplicity and the diversity of structures it produces. In two dimensions, D'Orsogna *et al.* [7] discovered rings, vortex-like swarms, and circular clumps in systems of self-propelled particles interacting via a Morse potential and showed how the swarm stability varies with swarm size. The Morse potential has been used to demonstrate the control of swarming vehicles [19], and in three dimensions was used to model systems of toroidal swarms whose translational motion was tuned with thermal noise [16].

Understanding swarm stability in thermal environments is particularly relevant in colloidal suspensions, where inter-particle interaction potentials can be qualitatively similar to the Morse potential. However, with four tunable parameters in the Morse potential and three independent thermodynamic parameters (e.g., volume, particle number, and temperature), there exists a large parameter space in which interesting swarms may exist. Because no predictive theory for swarm stability currently exists, computer simulations are the primary tool for predicting the conditions under which swarms are stable, and the structural and dynamical character of potentially achievable swarms.

Here we perform computer simulations of self-propelled Morse particles using graphics processing unit (GPU) optimized software [26] to explore the formation of three-dimensional stable swarms. Our simulation code allows for sampling of vast regions of parameter space where we observe swarms including tori, hollow shells, and two-dimensional rings. The "phase diagram" we report as a function of Morse potential parameters is surprisingly rich, with large regions where one swarm is stable over all others, but other regions where swarms are observed over only narrow parameter ranges. We observe notable deviations from equilibrium pattern formation in this far-from-equilibrium system, including assembly of competing swarms at a given state point, and we demonstrate that initial conditions can be chosen to bias the formation of one structure over another. We observe that thermal noise can influence the stability of one structure over another and we demonstrate how it can be used to reproducibly and repeatedly switch between different swarms. Beyond these new findings, our work highlights the need for efficient computational tools and predictive analytical techniques for the study of swarm formation, and demonstrates the precise control over swarm morphology that can be accomplished with a model system.

This paper is organized as follows. In Section 2 we describe the pairwise interactions and particle propulsions that define our model, the methods we employ to perform simulations on graphics processing units (GPUs), and the quantities we calculate to characterize swarms and their structure. In Section 3 we present the results of our extensive simulations. These results include a "phase diagram" that summarizes the swarms we find in the absence of thermal noise, two case studies for structural transitions induced by thermal noise, and an evaluation of swarm structure sensitivity to initial conditions. In Section 4 we discuss similarities and differences of the swarm self-assembly as compared to equilibrium self-assembly. In Section 5 we conclude with a summary of our work and highlight future extensions.

II. MODEL AND METHOD

A. Model

We consider particles interacting via the generalized two-body Morse potential [6]

$$U(r_{ij}) = C_a \exp\left(\frac{r_{ij}}{l_a}\right) - C_r \exp\left(\frac{r_{ij}}{l_r}\right) \tag{1}$$

where $r_{ij}$ is the distance between two particles $i$ and $j$, $C_a$ is the attraction strength, $C_r$ is the repulsion strength, $l_a$ is the attraction length scale, and $l_r$ is the repulsion length scale. These parameters can be chosen to model a wide range of interaction types, from purely repulsive and/or long-range attractive, to particles that can overlap but have an energetic barrier to doing so. To model the motion of self-propelled Morse particles in a thermal bath, we update particle positions using the Langevin equation of motion [27]

$$m_i \frac{\partial \vec{v}_i}{\partial t} = \vec{F}_i^C + \vec{F}_i^R + \vec{F}_i^D \quad . \tag{2}$$

Here $m_i$ and $v_i$ are the mass and velocity of particle $i$, $t$ is time, and $\vec{F}_i^C$, $\vec{F}_i^R$, and $\vec{F}_i^D$ represent the conservative, random, and drag forces, respectively. The conservative force between two particles is the usual negative gradient of the potential summed over all neighbors

$$\vec{F}_i^C = \sum_{i \neq j} -\nabla U(r_{ij}) \tag{3}$$

The random force and drag force are related through the fluctuation-dissipation theorem,

$$\langle \vec{F}_i^R \rangle = 0 \tag{4}$$

$$\langle \vec{F}_i^R(t) \vec{F}_j^R(t') \rangle = 6\gamma T^* \delta_{ij} \delta(t - t') \tag{5}$$

where $\gamma$ is proportional to fluid viscosity, $T^*$ is the dimensionless temperature, $\delta_{ij}$ is the Kronecker delta function and $\delta(t)$ is the Dirac delta function. We model particle self-propulsion as in Refs. [7, 16] with a modified drag force

$$\vec{F}_i^D = (\alpha - \gamma - \beta |\vec{v}_i|^2) \vec{v}_i \tag{6}$$

where $\alpha - \gamma$ determines the net propulsion strength, and $\beta$ determines the amplitude of the non-linear drag force. The propulsion and drag forces act parallel to a particle's velocity vector and define an optimal velocity $\vec{v}^* = \sqrt{\frac{(\alpha - \gamma)}{\beta}}$ towards which particles are driven.

B. Method

We implement the above model in HOOMD-blue, an open-source GPU-based Molecular Dynamics package developed by our group, which we have extended with packages for calculating and applying self-propulsion forces and the Morse potential via CUDA kernels executed on NVIDIA Tesla S1070 graphics cards [26]. The Morse potential (Eq. 1) is truncated and shifted to zero at $r_{ij} = 5\sigma$, which avoids the potential energy drift that can occur with an un-shifted potential. We employ generic units of distance $\sigma$, particle mass m and energy $\varepsilon$. The Morse potential energy at $r = 0$ is equal to $\varepsilon$ for $C_r = 2$, $C_a = 1$, $l_r = 1$, $l_a = 1$. The "temperature" of the system is a thermal energy related to real temperature by $T^* = k_B T_{real}$ in units of $\varepsilon$. The value of $k_B$ is uniquely determined by the choice of real units for energy, distance and mass. The time unit is derived from $\tau = \sqrt{\frac{m\sigma^2}{\varepsilon}}$, where $\sigma$ is a unit of interaction length. We update particle velocities and positions using the two-step velocity Verlet integration scheme using step sizes ranging from $dt = 0.001$ to $dt = 0.005$. Generally we find $dt = 0.005$ to be sufficiently small to converge and produce reproducible results, except for very low values of $C$ and large values of $l$, where artifacts can appear for $dt > 0.002$. With $N = 600$ point particles initialized randomly in a cubic simulation box with periodic boundary conditions and side length $V^{1/3} = L = 20\sigma$, a simulation of $2 \times 10^6$ time steps with $dt = 0.005$ requires about an hour. The 4000 simulations required to create our phase diagram were completed in three days, and would have required two to three months if performed with LAMMPS on a parallel CPU cluster [28,29]. We have reproduced the 2D results of Ref. [7] and the 3D results of Ref. [16] as validation of our GPU implementation.

A typical trial run begins with 600 particles initialized randomly within a cubic cell of edge length $5\sigma$ that is centered inside a larger cubic simulation box with edge length $20\sigma$ and periodic boundary conditions. Initializing particles within the $5\sigma$ cell allows all particles to interact from the initial time step, but avoids finite-size effects that could be imposed by having a simulation volume commensurate with the interaction potential length scale. Each component of particle velocities are initialized randomly from a Gaussian distribution with mean $T^*=1.0$ and standard deviation $\delta T^* =1.0$. Henceforth, when we refer to "random initial velocities" we mean they are drawn from this distribution. We set $\alpha =2.0$, $\beta =0.5$ and $\gamma =1.0$, and perform $2\times 10^6$ time steps with a step size of $dt =0.005$. Simulation runs that differ from these initial conditions will be noted and explained.

The simulation times of $2\times 10^6$ time steps are chosen to allow for sufficient sampling of steady state structures after transient swarms die out. We find that generally transients have disappeared after $1\times 10^6$ time steps. We define a steady state swarm for a simulation run if there exists a structure or collection of structures that persist over the final $1\times 10^6$ time steps with a well-defined average and standard deviation for their total energies. We distinguish among different swarms by calculating relative eigenvalues of their moment of inertia tensors and comparing them with the corresponding values that are characteristic to symmetric, circular structures such as spherical shells and rings. If the values match within an allowable tolerance of 10%, the structure is identified as a swarm.

To quantify correlations between the (typically circular) paths of particles traveling in a swarm, we define an alignment order

$$A = 2\max_{\forall i}\left(\frac{\sum_{j=1}^{N}\vec{M}_i \cdot \vec{M}_j}{\sum_{j=1}^{N}|\vec{M}_i||\vec{M}_j|} - 0.5\right) \tag{7}$$

where

$$\vec{M}_i = \vec{r}_i \times \vec{v}_i \tag{8}$$

is the angular momentum for particle $i$ traveling at a velocity $v_i$ located a distance $r_i$ away from its swarm center and $N$ is the total number of particles. The alignment order ranges from 0, which indicates no average correlation between particle angular momenta, to 1.0, which indicates all particles are traveling in the same (circular) path.

## III. RESULTS

### A. 3D Swarms

We perform 10 independent simulations at each of 400 different values of $\{l_r = 1, l_a, C_r = 1, C_a\}$, where $l = l_r/l_a$, $C = C_r/C_a$ and $T^* = 0$. In the range $0.1 \leq l \leq 2.0$ and $0.1 \leq C \leq 2.0$ we observe six distinguishable swarms: balls, shells, rings, spherical clumps, cylindrical clumps, and tori (Fig. 1). Given a single simulation run, the number of swarms observed after $2 \times 10^6$ time steps, their morphology, and their size can depend upon $l$, $C$, and, unlike in non-driven systems, details of the initial conditions. We summarize the swarm morphologies found as a function of $l$ and $C$ in the "phase diagram" of Fig. 2. For the ($l$, $C$) points at which all 10 of the runs resulted in one particular swarm, we have labeled the point with a symbol corresponding to that structure. In some regions of ($l$, $C$) space we find coexistence of shells and rings, or tori and balls. The area in the upper right of the phase diagram labeled "random" demonstrates no coalescence of particles into coherent structures. In the area labeled "mixture", located between the shell region and the random region, particles self-assemble into one or more small shells and random clusters. The random clusters typically consist of two to tens of particles that rotate in random trajectories about a translating center.

In the snowball structure (Fig. 1(a)), the particles organize into concentric spherical layers that minimize potential energy and travel in an identical linear trajectory. The snowball structure is only found at potentials that have short-range repulsion and long-range attraction. As the effective strength of the attraction increases, i.e. $C$ decreases and/or $l$ increases, we observe particles self-assemble into a hollow shell (Fig. 1(b)), a ring (Fig. 1(c)), clumps (Fig. 1(d,e)) or a torus (Fig. 1(f)). In each of these structures, the particles circulate about either a common axis or the structure's center of mass. In a hollow shell swarm, individual particles travel in circular paths on the surface of a sphere, although we observe no average correlation between particle orbits. In contrast, all particles share the same axis of rotation for the ring swarm. For the clumps particles organize into small clusters of particles that rotate about either a common center (spherical clumps, Fig.1(d)) or axis (cylindrical clumps, Fig.1(e)). Similar to a cylindrical clump, particles in a torus rotate about a common axis. Unlike the cylindrical clumps, the torus does not involve particles traveling in coherent clusters.

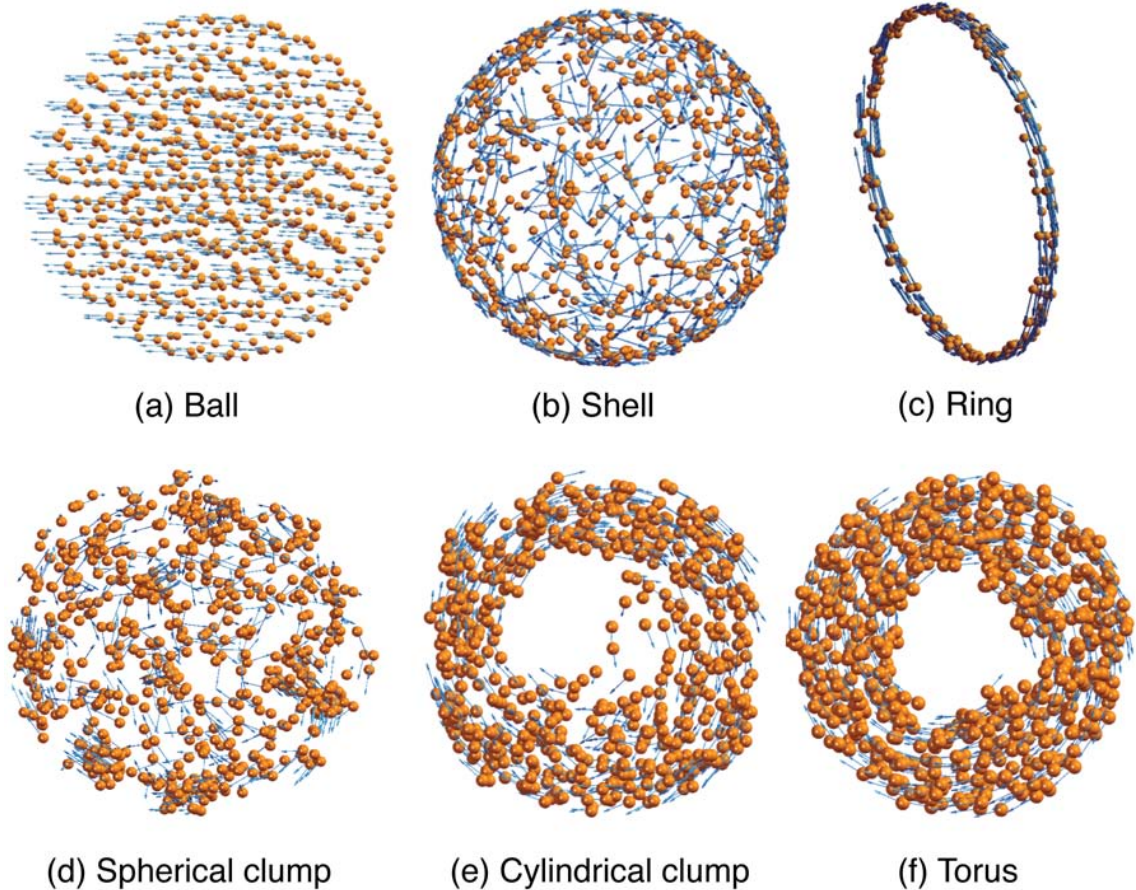

FIG. 1. Three-dimensional self-propelled swarms of Morse particles at $T^*=0$, $\alpha=2.0$, $\beta=0.5$, $\gamma=1.0$. All structures are observed in simulations that are initialized with 600 particles placed randomly within a $5\sigma$ cubic cell, after $2\times10^6$ time steps are performed within a $20\sigma$ cubic simulation box with periodic boundary conditions. The blue arrows indicate particle velocities. Note that the size of the orange spheres does not represent actual particle size, and that the radius of a swarm ranges from $0.06\sigma$ (ring) to $0.3\sigma$ (ball). (a) Ball with translational velocity $\vec{v}^*$ composed of concentric spherical layers of particles. (b) Stationary hollow shell composed of particles that travel in circular orbits. (c) Stationary ring with all particles traveling in nearly the same circular orbit. (d) Stationary spherical swarm with clumps of particles that travel in circular orbits. (e) Stationary cylindrical swarm composed of clumps of particles traveling in nearly the same circular orbit. (f) Stationary torus composed of particles that travel in circular orbits of different radii, but share an axis of rotation.

At higher values of $C$ and $l \geq 1$, mixtures of structures (denoted by "Mixture" in Fig. 2) and gas-like collections of randomly distributed particles (denoted by "Random" in Fig. 2) are formed. In the Mixture region, multiple clusters, including shells and small balls, co-exist in the simulation box. The interaction profile for the Mixture region is characterized by a maximum in potential energy that creates an energetic barrier to particle aggregation. As $C$ is increased in the Mixture region, the Morse potential maximum shifts towards $r = 0$, which makes it increasingly difficult for large collections of particles to aggregate. When $C$ becomes sufficiently large, particles are no longer able to agglomerate and a random "gas" of particles is observed.

Some of the structures we report here are analogues of the two-dimensional swarms predicted in Ref. [6], but others are surprising or have no 2D analogues. For example, it is natural to think that the 3D analogue of a 2D ring is a hollow spherical shell, or that circular clumps of particles in a circular swarm in 2D might be analogous to spherical clumps of particles swarming in a torus in 3D. However, it is not obvious that we should observe 2D rings in our 3D simulations and that they should be stable across many ($l, C$) values. We observe coexistence of shells and rings at $l = 0.9$ and $C = 0.9$, which is interesting because swarm coexistence is not reported in 2D systems. In a departure from equilibrium statistical mechanics, we find that in the regions of swarm coexistence the probability of forming one swarm can be biased by choice of initial conditions. We discuss in detail some characteristics of initial condition sensitivity and swarm coexistence in the following section.

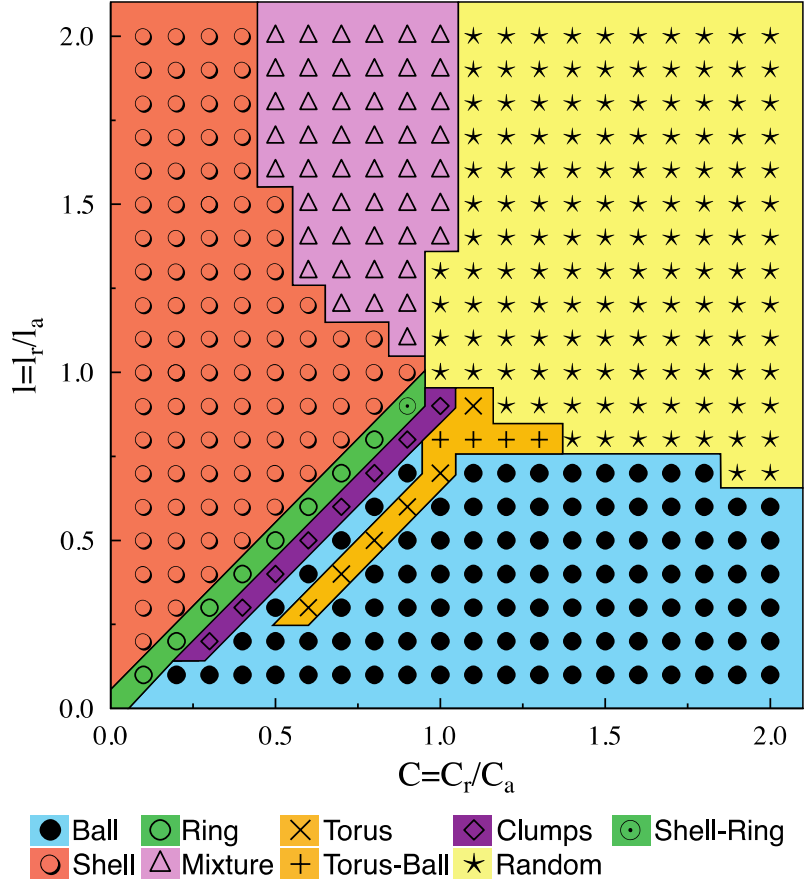

FIG. 2. "Phase diagram" for 3D stable structures assembled at values of $0.1 \leq l \leq 2.0$ and $0.1 \leq C \leq 2.0$, $T^* = 0$, $\alpha = 2.0$, $\beta = 0.5$, and $\gamma = 1.0$. The value $C = C_r/C_a$ is the ratio of repulsion to attraction strength and $l = l_r/l_a$ is the ratio of the repulsion and attraction length scales. In the upper right quadrant ($l > 1$, $C > 1$) repulsion dominates the potential, preventing organized clusters from assembling. In the lower right quadrant ($l < 1$, $C > 1$) the repulsion is short-ranged and the attraction is long-ranged, resulting in translating spherical balls. In the lower left quadrant ($l < 1$, $C < 1$) attraction dominates the potential and the phase diagram is most complex, with shells, rings, tori, balls, and clumps all stabilized within narrow ranges of $l$ and $C$. In the top left quadrant ($l > 1$, $C < 1$) particles are attractive, but the repulsive length scale dominates, giving rise to hollow shells.

B. Initial Conditions

The phase diagram in Fig. 2 summarizes the types of swarms predicted to self-assemble from particles initialized randomly in a $5\sigma$ cubic cell contained within a much larger simulation box, and has regions where multiple structures might assemble from a random initial configuration. In this subsection we consider three regions of the phase diagram and investigate ways in which initial conditions can be chosen to bias swarm formation; in all cases we choose $T^* = 0$.

We first consider shell formation at ($l = 1.7$, $C = 1.2$), where we observe no ordered structure assembling from particles initialized randomly within a $5\sigma$ cube. If we instead initialize particles as a small shell that fits within the attraction region of the potential, we observe that the shell shrinks or expands, depending on its initial size, to another shell that becomes stable for the remainder of $1\times10^6$ time step simulations. The necessary condition for shell formation is that particles are initialized within the attraction region and have sufficiently low kinetic energy that they will not escape to the repulsive regime of the Morse potential. Indeed, we also find shells by initializing particles on a single point with random initial velocities. Moreover, shells can also be formed at other state points in the Mixture and Random regions. By initializing the system as a shell, we find shells form within the region ($l = 1.7$, $C \leq 1.5$).

Next we consider ($l = 0.8$, $C = 1.0$), where the ball and torus coexist in our simulations. We perform simulations initialized as a ball, as a cylinder, and randomly distributed within a $5\sigma$ cube, with 100 independent runs with different random initial velocities. At these parameters, we find the probability of observing a toroidal swarm to be 6% for all three initial conditions. In all cases that the torus does not form, a spherical ball forms instead and translates through the simulation box.

Finally, we consider ring and shell coexistence at ($l = 0.5$, $C = 0.5$), We note that whenever we observe a ring form, we also observe that it is preceded by a hollow shell which then transforms into a ring. The shell-to-ring transition occurs over $4\times10^4$ time steps, and can occur as many as $4\times10^4$ time steps after the shell has self-assembled. We perform simulations in which particle positions are initialized as a shell, sphere, cylinder, and randomly distributed within a $5\sigma$ cube to investigate sensitivity to initial conditions at this state point. For each initial condition we perform 100 runs with random initial velocities and record the structure observed after $1\times 10^{6}$ time steps (Fig. 3). When initialized as a shell, only 2% of swarms transition into rings, a substantially lower percentage than the 98% of swarms that form rings from spatially randomized initial conditions, all of which pass through a hollow shell transition state. For each of the 100 runs initialized as a cylinder, all transition into rings. For the runs initialized as balls, 12% transition into rings.

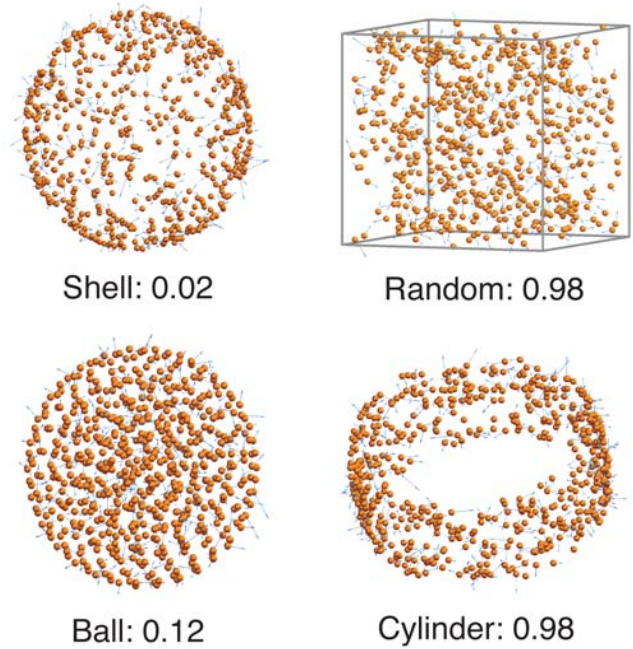

FIG. 3. The probabilities of generating a ring for different initial conditions. One hundred independent runs with different pseudorandom number generator seeds are performed for the indicated initial condition with $T^{*}= 0$ at $\alpha = 1$, $\beta =0.5$, $C_r =0.5$, $l_r =0.5$, $C_a = 1$, $l_a = 1$, $N = 600$ and with different random initial velocities.

The fact that ring formation is always preceded by shell formation is worthy of further investigation. Why do some shells collapse into rings, while others remain spherical for the entirety ($1\times 10^{7}$ time steps) of a simulation? We initialize 1000 shells with random initial velocities and random particle positions within the shells and run for $1\times 10^{6}$ time steps. For each simulation run we observe either a shell or a ring as the steady-state configuration and we find a strong correlation between ring formation and initial alignment order, as defined in Eq. 7 (Fig. 4).

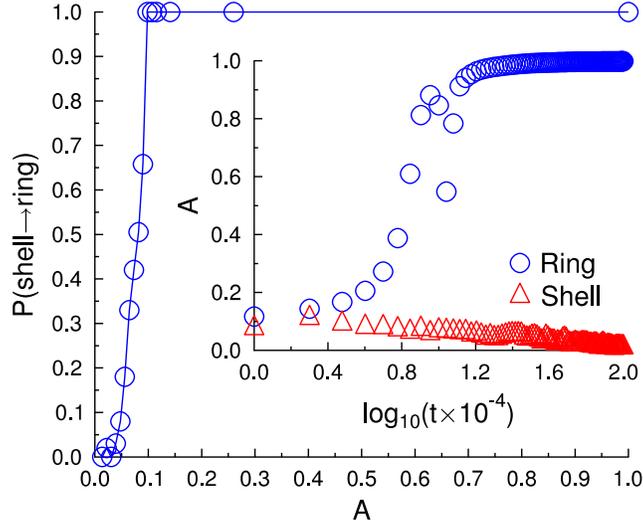

FIG. 4. Probability of a random shell transforming into a ring as a function of alignment order $A$ (Eq. 7) at $T^*=0$. The blue line is drawn as a guide to the eye. The insets are representative alignment order trajectories for swarms that form a shell (red) and ring (blue). The figure shows probability of shell-ring transformation as a function of alignment order at $T^* = 0$. $\alpha = 1$, $\beta = 0.5$, $\gamma = 1.0$, $C_r = 0.5$, $l_r = 0.5$, $C_a = 1$, $l_a = 1$, $N = 600$.

The alignment order for a swarm is a function of both particle velocities and positions, the distributions for both of which are influenced by temperature. Using the same 1000 spherical shell swarms as above with $T^*=0.008$, we find a negligible difference in the probability distribution of alignment order compared with $T^*=0$. We also find that the distribution shape is not correlated with shell roughness, which is a measure of how much the particles deviate from the mean radius of the shell.

### C. Noise

In this section we perform simulations at $T^* > 0$ at two state points to investigate how thermal fluctuations can influence the structures assembled in Fig. 2. We investigate the point ($l = 0.5$, $C = 0.5$) where rings and shells coexist at $T^* = 0$, and the point ($l = 0.8$, $C = 1.0$) where tori and balls coexist. As temperature is increased for both of these cases, we observe that thermal noise provides a stabilizing effect for one structure over all others, provided the noise is not too large. If the noise is too large, no ordered structure can be maintained.

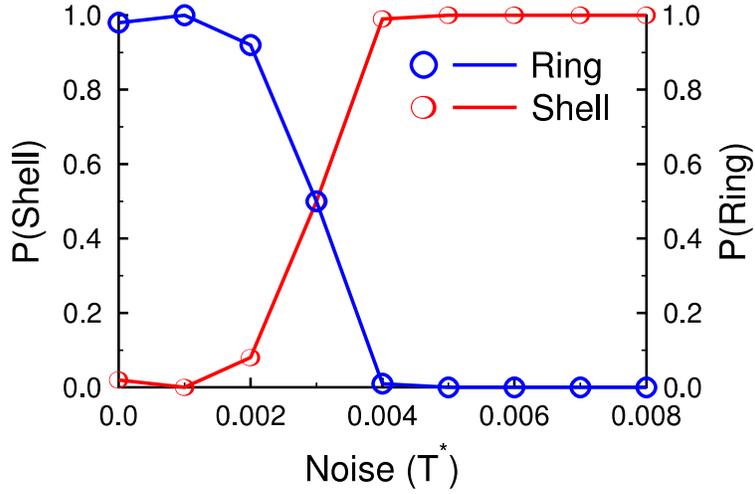

FIG. 5. Probability of shell (red) and ring (blue) formation as a function of $T^*$ at $\alpha = 2$, $\gamma = 1$, $\beta = 0.5$, $C_r = 0.5$, $l_r = 0.5$, $C_a = 1$, $l_a = 1$, $N = 600$. 100 independent runs are performed at each noise level, with particles initially randomized in a $5\sigma$ cubic box in a $20\sigma$ cubic simulation box and are evolved for $1\times 10^6$ time steps.

At the shell/ring coexistence point ($l = 0.5$, $C = 0.5$) we perform simulations at $T^*=0$ to $T^*=0.008$ at increments of $\delta T^*=0.001$. For each of these temperatures we perform 100 independent simulations initializing $N=600$ particles randomly in a $5\sigma$ cube within a $20\sigma$ cubic simulation box with periodic boundary conditions and allow the system to advance for $1\times 10^6$ time steps. At these temperatures we find that hollow shells and rings self-assemble, but for $T^* > 0$ there are two significant changes. First, we find that the probability of assembling a ring decreases as $T^*$ increases (Fig. 5), undergoing a sharp transition at $T^*=0.003$, above which shell self-assembly is enhanced.

The second change we observe at $T^* > 0$ is that the ring structures are no longer flat, but cylindrical. We perform additional simulations initialized from a ring configuration and ramp temperature from $T^* = 0$ to $T^* = 0.008$ by a step increase in temperature of $\delta T^* = 0.001$ every $1\times 10^6$ time steps, followed by a reverse ramping back to $T^* = 0$, for a total of

$16 \times 10^6$ time steps. We perform these simulations for swarms of size $N = 100$, $N = 200$, $N = 400$, and $N = 600$ and find that cylinder height increases linearly with $T^*$ from $T^* = 0.001$ to $T^* = 0.007$ (Fig. 6). Furthermore, we find that by normalizing the cylinder height by the cylinder diameter, this trend is independent of swarm size (Fig. 6). When $T^* > 0.007$, cylinders transition into hollow shells. This transition can be used as part of a reversible sequence of steps to convert rings into cylinders into shells and back into rings (Fig. 7). We find that by dropping the temperature from $T^* = 0.008$ to $T^* < 0.007$, a shell will transform into a ring or cylinder (depending on the temperature) if its instantaneous alignment order is sufficiently large (Fig. 4).

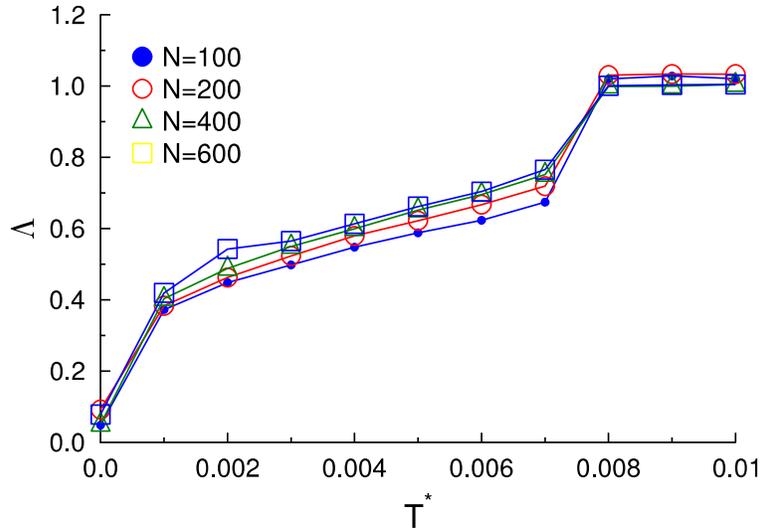

FIG. 6. Normalized cylinder height ($\Lambda$) as a function of dimensionless temperature. Particles are initialized as a ring with Gaussian velocity distributions, $\alpha = 2$, $\gamma = 1$, $\beta = 0.5$, $C_r = 0.5$, $l_r = 0.5$, $C_a = 1$, and $l_a = 1$. $\Lambda$ are averaged over 10 simulation snapshots and over 10 independent simulation runs, where the snapshots are the last 10 at $1 \times 10^4$ time step increments.

At ($C_r = 0.9$, $l_r = 0.6$, $C_a = 1$, $l_a = 1$), a parameter combination at which tori are found, we find similar noise sensitivity. We gradually increase the noise intensity from 0 to 0.08 and then reduce the intensity back to zero with a change of $\delta T^* = 0.01$ every 1 million time steps. We find that when $T^* > 0.03$ the torus transforms to a bumpy sphere, which diffuses through the simulation box. As the noise intensity is reduced to zero, the bumpy sphere transitions back to a stationary torus.

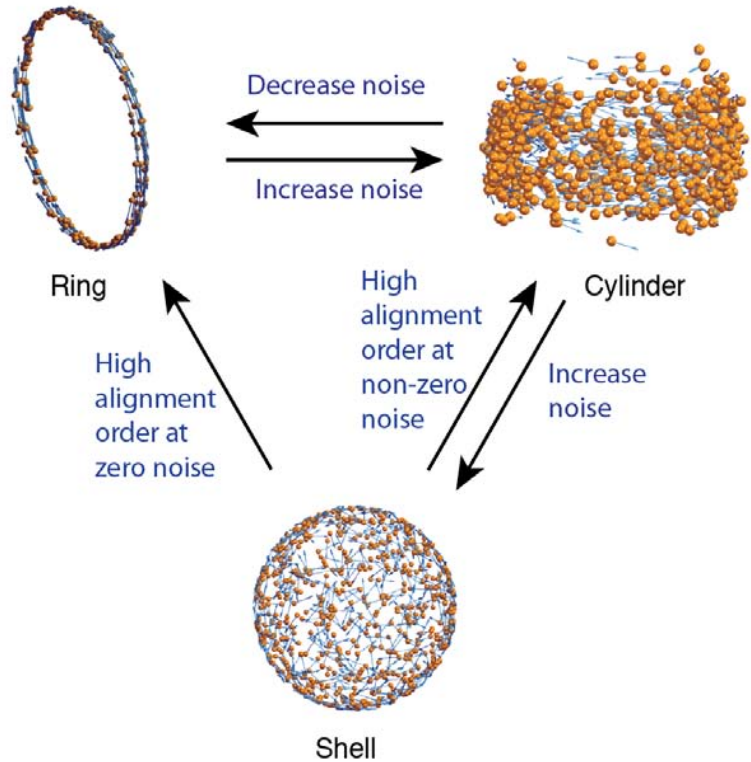

FIG. 7. Mechanisms of structural transitions among shells, rings, and cylinders. Increasing the temperature of a ring causes it to transition into a cylinder of increasing height. At large enough $T^*$, cylinders transition into hollow shells. The transformation of a hollow shell into a cylinder or ring requires not only that the temperature be decreased below $T^*=0.007$, but also that an instantaneous fluctuation in alignment order is sufficiently large to initiate the transition.

IV. DISCUSSION

Swarm formation in self-propelled particles can be considered as a perturbation to equilibrium self-assembly. In the limit that $\alpha \to 0$ and $\beta \to 0$, Eq. 2 becomes the traditional Langevin equation modeling Brownian motion, from which equilibrium distributions of configurations can easily be sampled. From the simulations sampled here with nonzero $\alpha$ and $\beta$, we observe the formation of swarms over a wide range of Morse potential shapes, but for which the variational principle controlling structure formation is not simply minimization of a traditional (e.g. Helmholtz) free energy. Although aspects of free energy minimization are apparent in the swarm structures we observe, including mechanical stability and entropy maximization, additional theory is needed to predict whether one structure is stable relative to another.

The two most relevant driving forces towards steady-state swarm formation are the balance of conservative forces from neighboring particles and the simultaneous achievement of the optimal velocity $\vec{v}^* = \sqrt{\frac{(\alpha-\gamma)}{\beta}}$. In the case that all particles achieve identical, parallel $\vec{v}^*$, we find that the patterns formed minimize potential energy in a translating frame of reference. The ball swarm (Fig. 1(a)) is an example of such a solution, stable across a large region of ($l$, $C$), and the analogous 2D swarm has been observed before [7, 16, 17]. The other swarm solution we observe features particles traveling at $\vec{v}^*$ on average, but in a circular swarm in which centripetal forces are balanced with the conservative forces arising from the Morse potential. Rings, shells, cylinders, circular and cylindrical clumps, and tori are examples of this second type of solution. Every steady state swarm we observe in our simulations falls into one of these two categories.

That mechanical stability and achievement of $\vec{v}^*$ are driving forces may seem trivial, but the degree to which one matters more than the other is not obvious in the case of ring and shell coexistence. The consistent observation that rings are preceded by hollow shells at $T^* < 0.007$ suggests that rings are more stable than shells, which is corroborated by the fact we never observe the reverse transition. The fundamental difference between the ring and the shell is that the fluctuations about $\vec{v}^*$ are smaller in the ring, suggesting that the degree to which particles maintain $\vec{v}^*$ matters for stability. The fact that there is a temperature above which rings transition to shells is consistent with what we would expect from equilibrium thermodynamics; with sufficiently large thermal fluctuations, configurations with more ways of satisfying the driving forces (higher entropy) are preferred over those that require perfect force balance.

These similarities with equilibrium self-assembly suggest that there might be some form of an "extended" free energy that can be written for self-propelled swarms that provides predictive capabilities. Previous work by Schweitzer *et al.* [15] provides a promising direction for such a development. The free energy would be extended in the sense that it adds additional non-equilibrium driving forces to a well-known equilibrium ensemble free energy. For the systems we study here, it seems natural that this extended free energy should include $\alpha$, $\beta$, and $\gamma$, as these parameters provide the deviation from a standard equilibrium simulation.

Derivations for the functional form and demonstration of such a free energy are beyond the scope of the current work, but are an exciting challenge with important implications for theories of nonequilibrium self-assembly. Regions of parameter space in which structures coexist provide test-cases for points at which extended free energies should equate, therefore models such as the propelled Morse particles studied here are ideal candidates for theory development.

## V. CONCLUSIONS

We have performed extensive GPU-enabled simulations of self-propelled swarming particles, characterizing the structures that can be assembled for just one functional form of interparticle interactions. The diversity of the swarms stabilized in this system, their coexistence regions, and their sensitivity to thermal fluctuations (noise) highlights the need for efficient computer simulations with which parameter space can be explored and detailed experiments performed. We observe behaviors that are marked deviations from equilibrium self-assembly, including swarm coexistence and sensitivity to initial conditions, but our results suggest that modifications to equilibrium statistical mechanics may have predictive capabilities. We propose a way forward, combining the swarms found in regions of coexistence as test cases for extended free energy development. Finally, the diversity of structures we observe here and the demonstrated control over their morphology may have immediate implications in the exploitation of swarms of practical interest.

## VI. ACKNOWLEDGMENTS

This work was supported by the Non-Equilibrium Energy Research Center (NERC), an Energy Frontier Research Center funded by the U.S. Department of Energy, Office of Science, Office of Basic Energy Sciences under Award Number DE-SC0000989. EJ received support from the James S. McDonnell Foundation 21st Century Science Research Award/Studying Complex Systems, grant no. 220020139 and from a National Defense Science and Engineering Graduate (NDSEG) Fellowship, 32 CFR 168a. NN also acknowledges the Vietnam Education Foundation. We thank Carolyn Phillips and Trung Dac Nguyen for helpful discussions and assistance, and Daphne Klotsa for comments on this manuscript.